\documentclass[12pt,a4paper]{article}
\usepackage{graphicx}
\begin{document}
\textwidth=135mm
 \textheight=200mm
\begin{center}
{\bfseries Pion-proton correlations and asymmetry measurement in Au+Au collisions at $\sqrt{s_{NN}}=200$ $GeV$ data
\footnote{{\small This work has been partially supported by the Polish Ministry of Science and Higher Education under grant No N N202 173235.}}
\footnote{{\small Talk at the VI Workshop of Particle Correlations and Femtoscopy; Kiev, September 14-18th 2010}}}

\vskip 5mm
M. Zawisza$^{\dag}$ \\
For the STAR Collaboration
\vskip 5mm
{\small {\it $^\dag$ Faculty of Physics, Warsaw University of Technology, Koszykowa 75, 00-662 Warsaw, Poland}}
\\
\end{center}
\vskip 5mm
\centerline{\bf Abstract}
Correlations between non-identical particles at small relative
velocity probe asymmetries in the average space-time emission points at
freeze-out. The origin of such asymmetries may be from long-lived
resonances, bulk collective effects, or differences in the freeze-out
scenario for the different particle species.
STAR has extracted pion-proton correlation functions from a
dataset of Au+Au collisions at $\sqrt{s_{NN}}=200$ $GeV$.
We present correlation functions in the spherical harmonic
decomposition representation, for different centralities and for 
different combinations of pions and (anti-)protons.
\vskip 10mm
\section{\label{sec:intro}Introduction}
Femtoscopy is the primary technique that provides possibility of measuring space-time quantities of the order of $10^{-15}$m and $10^{-23}$s.
It allows for investigation of evolution of nuclear matter created in heavy ion collisions.
Two charged particles with small relative velocity emitted from source interact through strong and Coulomb interactions called final state interactions (FSI) which give correlation between them.
Correlations of non-identical particles are sensitive to space-time asymmetry in emission of particles~\cite{asym:llen}\cite{asym:vlpx}.
By measuring them we can investigate which sort of particles are emitted earlier or later and what is the relative shift between average emission points.

Information about size and space-time asymmetry of the source can be extracted from correlation functions calculated with respect to the sign of the $out$ (along the velocity of the pair in transverse plane), $long$ (along the beam axis) and $side$ (in transverse plane, perpendicular to the other two) components of the vector $k^*$ (momentum of first particle in Pair Rest Frame). Such preliminary pion-proton correlation functions measured by STAR were presented in~\cite{mzappb}. However such information is most efficiently encoded in a spherical harmonics representation of the correlation function \cite{shzibimike}. In case of limited momentum acceptance of the detector or big difference in mass between particles, when some orientation in space of the $k^{*}$ vector are not registered by the detector, calculation of correlation function directly in spherical harmonics is the most effective method~\cite{shkisielbrown}.
Functions calculated directly in spherical harmonics are less sensitive to momentum acceptance limitations than functions calculated from 3-dimensional correlation function.

\section{Experimental procedure}
This work presents analysis of data taken in 2004 by the STAR experiment. Au+Au collisions at $\sqrt{s_{NN}}=200$ $GeV$ are analyzed in 3 different centrality bins: 0-10\% - central events, 10-30\% - intermediate events and 30-50\% - mid-central events. The particle identification is done in Time Projection Chamber (TPC). Both pions and protons are selected based on $N_\sigma=[-3.0,3.0]$ value which describes distance from mean of the Gaussian parametrization of the $dE/dx$ curve. Particles are selected from the mid-rapidity region $|y|<0.7$. Transverse momentum of the pion is set within the range of $p_T=[0.1,0.6]$ $GeV/c$ and for proton $p_T=[0.4,1.25]$ $GeV/c$.

Pion-proton pairs are constructed from close velocity particles. Due to significant mass difference between pions and protons, pion with momentum 0.18 $GeV/c$ requires proton with momentum 1.2 $GeV/c$. This shows that this analysis reaches the limits of the particle identification capabilities of the STAR TPC detector. At the level of pair selection, pairs of tracks which share more than 10\% of the hits in the detector are removed from the analysis (the so called hit merging effect). We also eliminate correlated electron-positron pairs which come from gamma conversion (based on topological cut) and non pion-proton pairs (based on probability).

Background is constructed only from events with similar characteristics thus events are divided into bins with respect to z-vertex position - 15 bins, event multiplicity - 6 bins and mean transverse momentum of the registered particles - 3 bins.

To obtain experimental correlation function we construct distribution of the correlated pairs (coming from the same event) and distribution of the uncorrelated pairs (coming from different events). Distributions are directly binned in spherical harmonics. Detailed explanation of the formalism can be found in~\cite{shkisielbrown}.

In our analysis $C^0_0(k^*)$ and $Re C^1_1(k^*)$ components are analyzed. $C^0_0$ represents overall size of the system. $Re C^1_1$ component represents space-time asymmetry observed in the $out$ direction.

Constructed functions are corrected on purity which is defined as a product of PID probability of the particles in the pair and fraction of primary pairs.

\section{Results}
\begin{figure}[h]
\begin{center}
\includegraphics[width=0.9\textwidth]{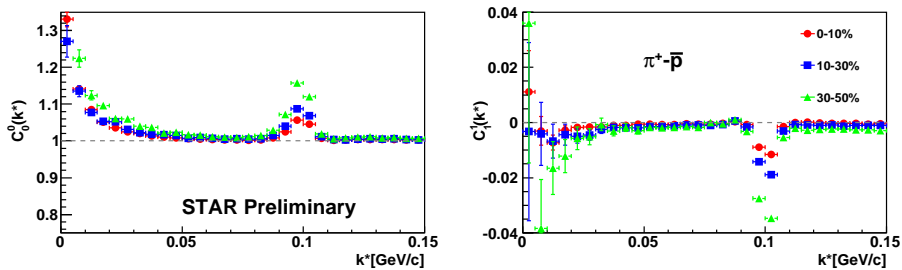}
\caption{\label{shpcpm}$\pi^+-\bar p$ correlation functions. Left panel - $C^0_0$ component, right panel - $ReC^1_1$ component. Red points - central 0-10\% events, blue points - intermediate 10-30\% events and blue points - 30-50\% mid-central events.}
\label{cfpippm}
\end{center}
\end{figure}

\begin{figure}[h]
\begin{center}
\includegraphics[width=0.9\textwidth]{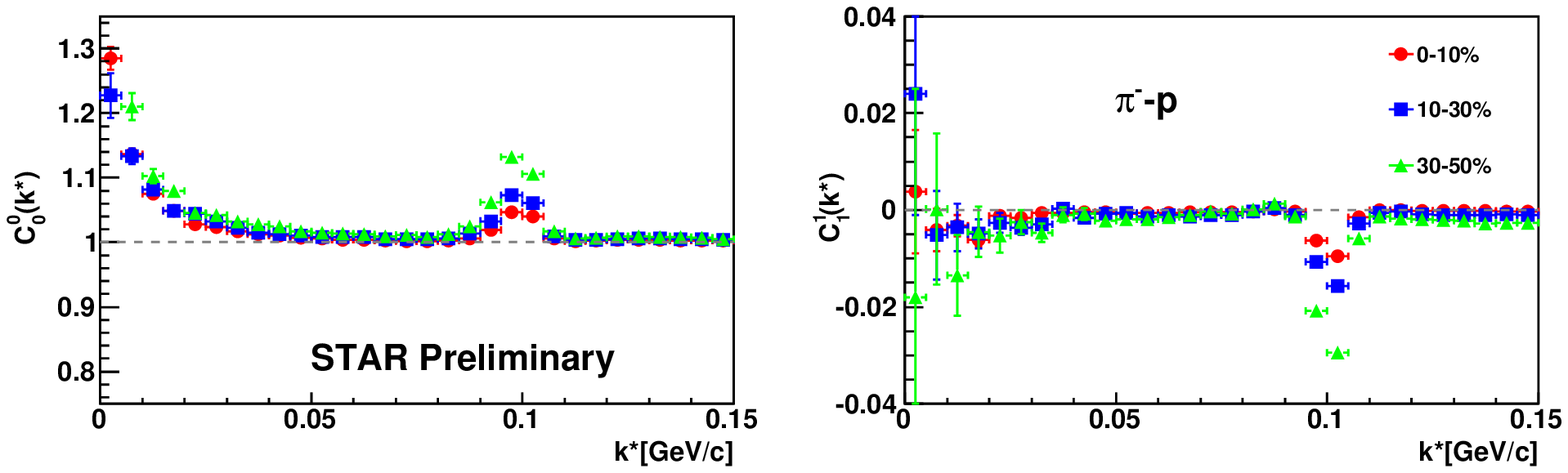}
\caption{\label{shpcmp}$\pi^--p$ correlation functions. Left panel - $C^0_0$ component, right panel - $ReC^1_1$ component. Red points - central 0-10\% events, blue points - intermediate 10-30\% events and blue points - 30-50\% mid-central events.}
\label{cfpimpp}
\end{center}
\end{figure}

\begin{figure}[h]
\begin{center}
\includegraphics[width=0.85\textwidth]{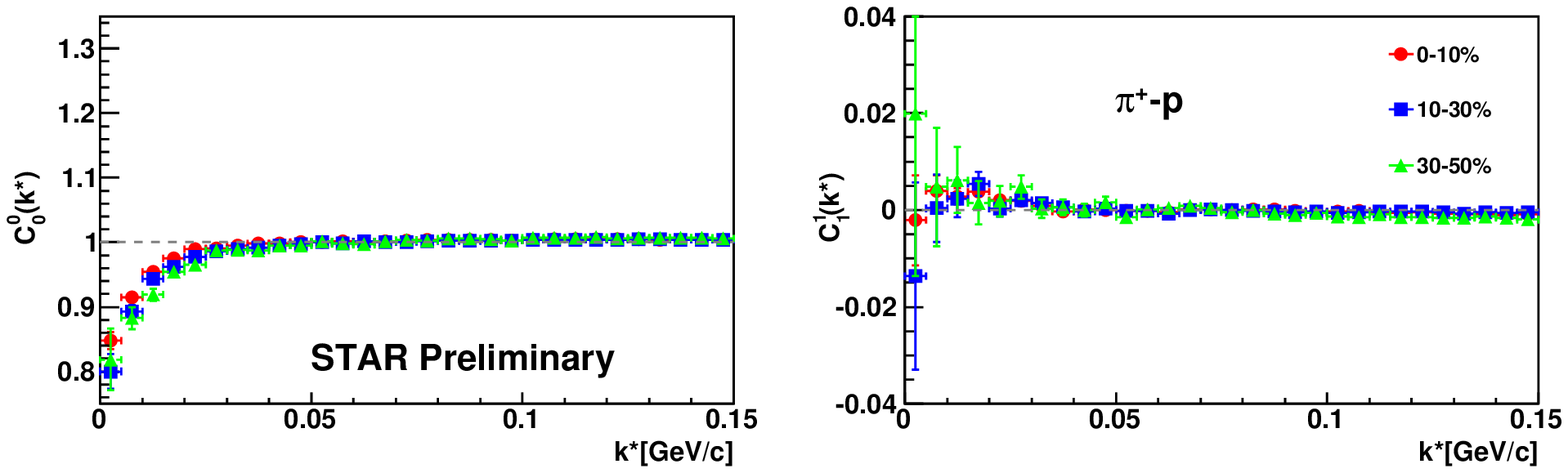}
\includegraphics[width=0.85\textwidth]{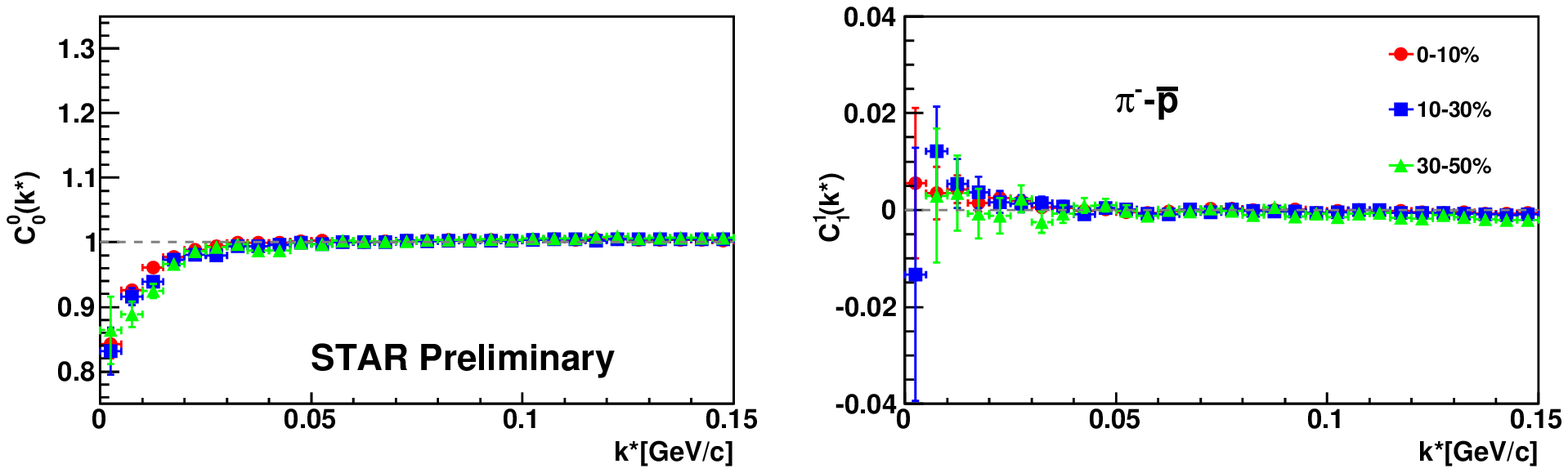}
\caption{\label{shpcppmm}$\pi^+-p$ (up) and $\pi^--\bar p$ (bottom) correlation functions. Red points - central events, blue points - intermediate events and blue points - mid-central events.}
\label{cfpipppmm}
\end{center}
\end{figure}

Figures \ref{cfpippm} and \ref{cfpimpp} show $C^0_0$ and $C^1_1$ components of the correlation function for the $\pi^+-\bar p$ and $\pi^--p$ pairs correspondingly calculated for Au+Au collisions at $\sqrt{s_{NN}}=200$ $GeV$. Red dots show central data, blue squares are for intermediate events and green triangles are for mid-central events. Peak around $k^*=0.1$ $GeV/c$ reflects correlated products of the (anti)lambda decay.
Lambda peaks do not overlap with femtoscopic correlation effect.
$C^0_0$ functions show positive correlation as the interaction between unlike sign particles is repulsive. $Re C^1_1$ functions show small deviation from zero for small values of $k^*$ which suggests that pions and protons are not emitted from the same regions of the source or are not emitted at the same time. Figure \ref{cfpipppmm} presents functions for like sign pairs where correlation effect is negative as a Coulomb interaction is attractive. $C_1^1$ components for like sign pairs also show asymmetry.

\begin{figure}[h]
\begin{center}
\includegraphics[width=0.5\textwidth]{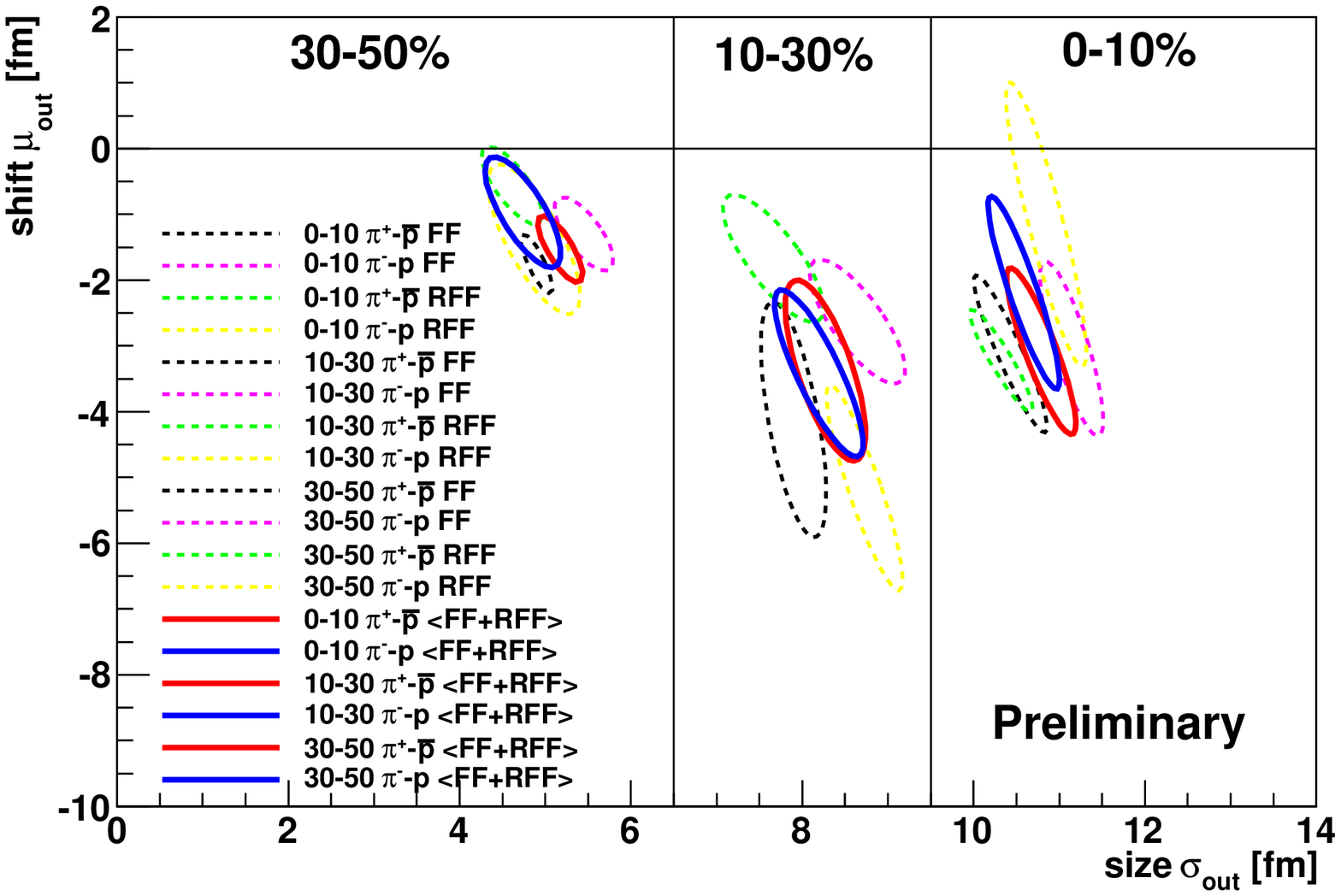}
\includegraphics[width=0.5\textwidth]{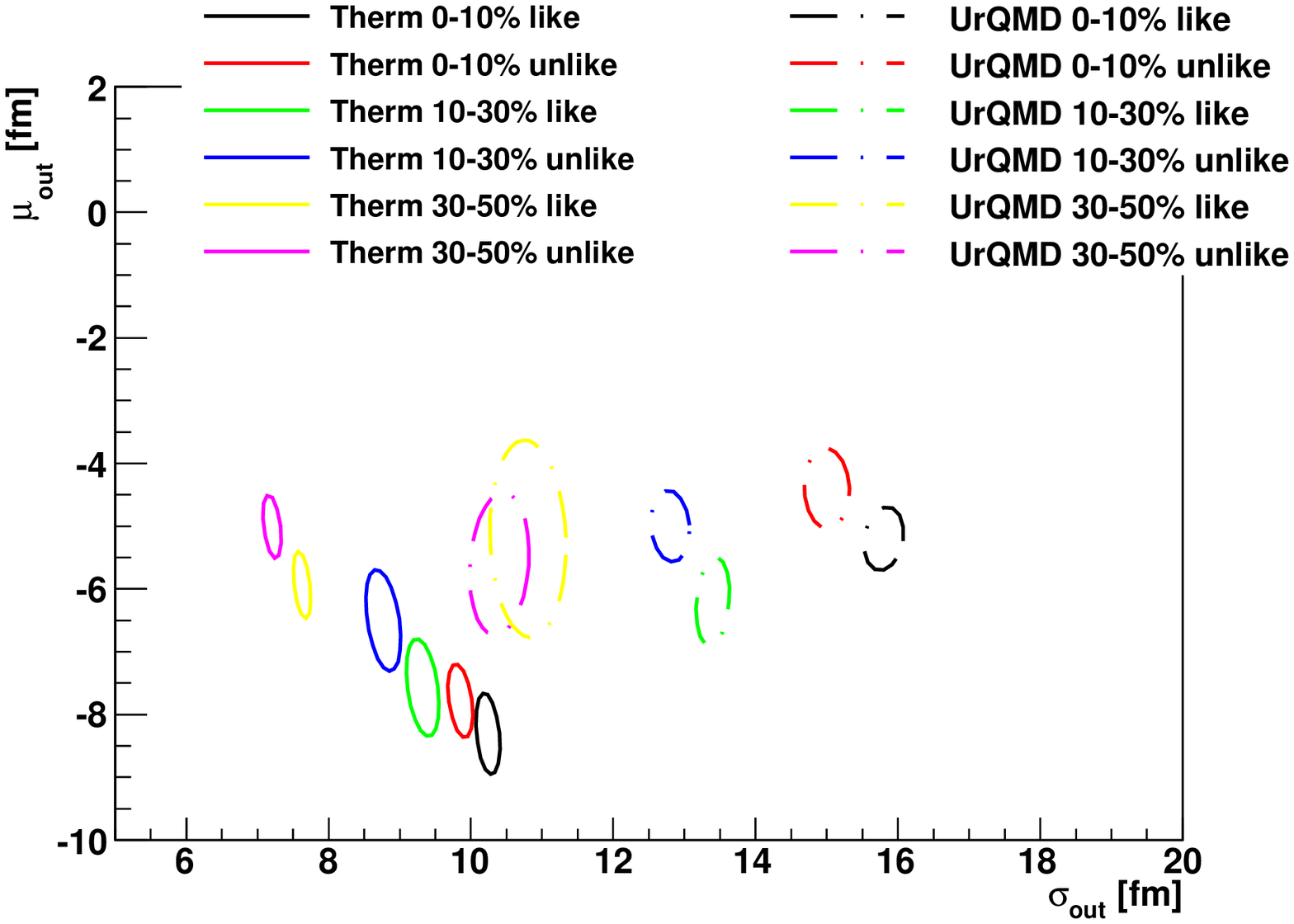}
\end{center}
\caption{Ellipses of covariances show the 3D Gaussian fit results of the experimental (left) and theoretical (right) correlation functions. Experimental results are for two orientations of the magnetic field in the detector (dash-dot line). Continuous lines show averaged results.}
\label{fit}
\end{figure}

In the fit procedure 3-dimensional Gaussian profile of the source is assumed. Set of correlation functions with various parameters of the source profile is calculated\cite{corrfit}. Then around the best $\chi^2$ value an ellipse of covariance is calculated. Figure~\ref{fit} shows results for experimental data as well as results obtained for Therminator+Lhyquid\cite{therm} and UrQMD v3.3\cite{urqmd} data. Experimental results obtained for unlike-sign pairs show centrality dependence of the size ($\sigma_{out}$) and asymmetry ($\mu_{out}$) of the source. Model study is done for both like-sign and unlike-sign pairs. In UrQMD data observed asymmetry seems to be constant for investigated centralities.

\begin{figure}[h]
\begin{center}
\includegraphics[width=0.45\textwidth]{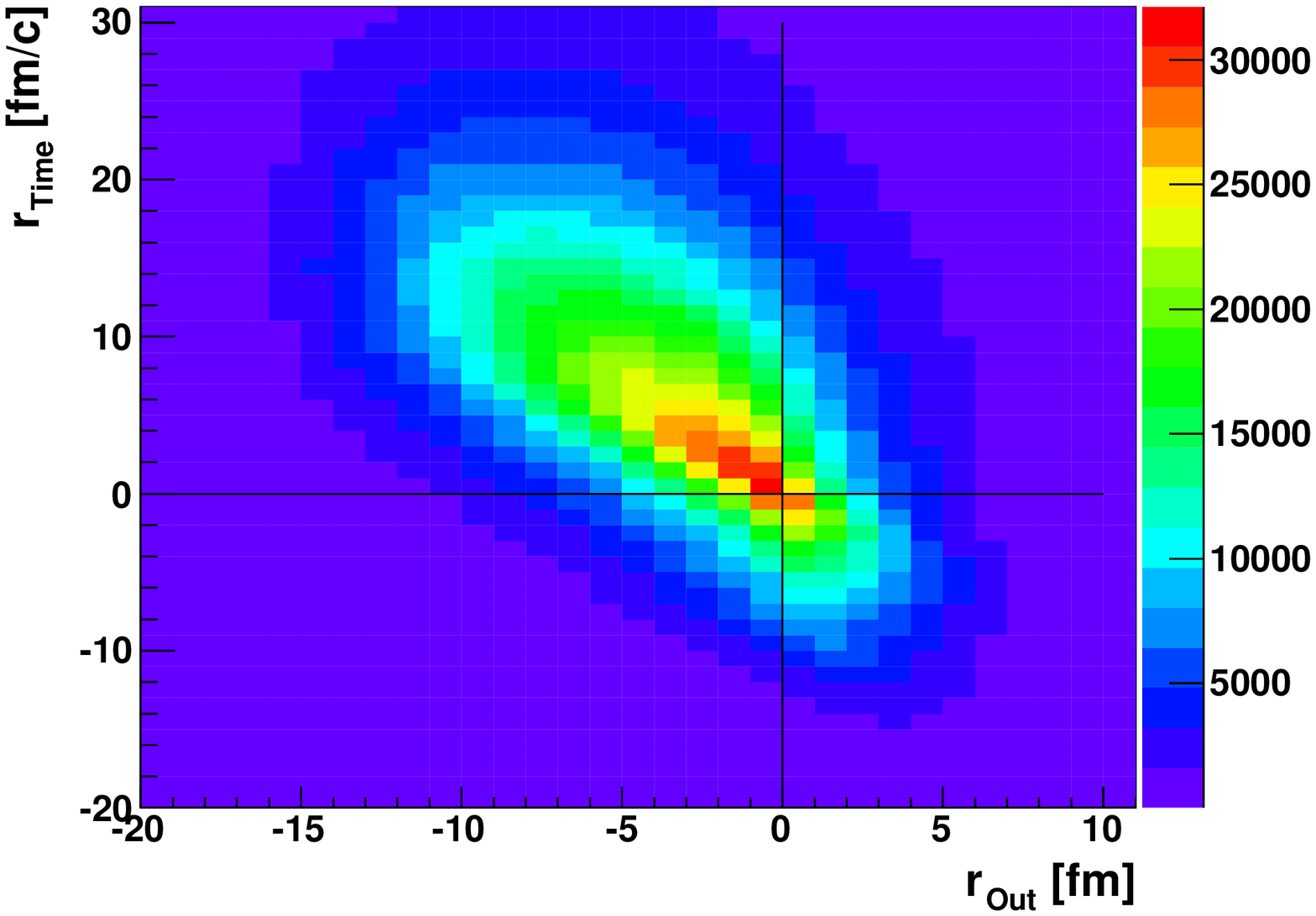}
\includegraphics[width=0.45\textwidth]{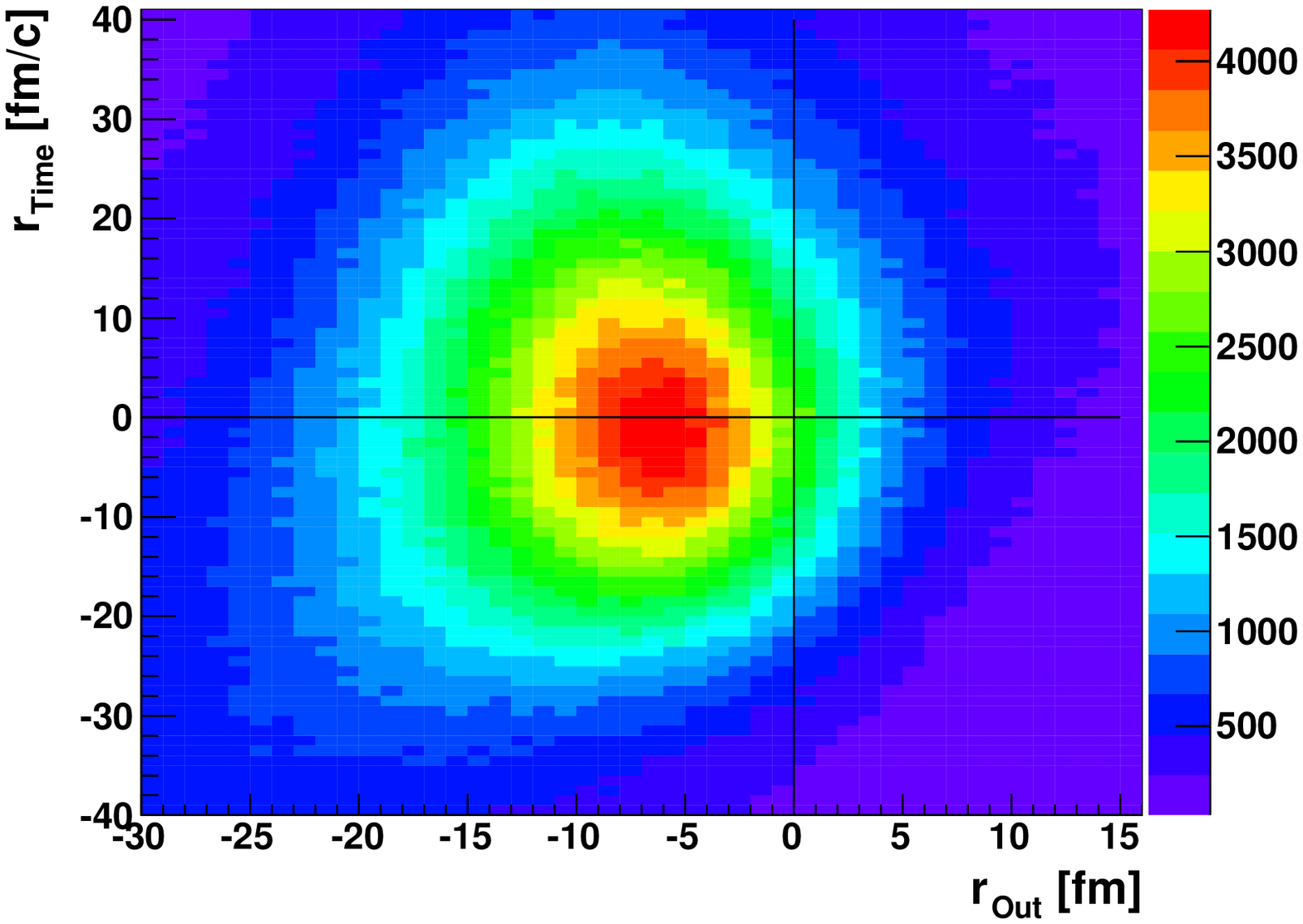}
\end{center}
\caption{Distribution of the emission time difference versus separation of the emission points in Therminator (left) and UrQMD (right).}
\label{outtime}
\end{figure}

Experimental procedure allow only for measuring total asymmetry, more detailed study with models gives more detailed information. Figure~\ref{outtime} shows distribution of the emission time difference versus separation of the emission points calculated in Therminator and UrQMD. Positive values of $r_{Time}$ corresponds to pion emitted later than proton. Such scenario dominates in Therminator data. Negative values of $r_{out}$ means that emission point of proton is shifted outward (to the edge of the source) with respect to the emission point of pion.
\section{Summary}
Analysis presented here shows that in system created in Au+Au collisions at $\sqrt{s_{NN}}=200$ $GeV$ average emission points of pions and protons are not the same. Observed asymmetry depends on centrality and is correlated with the size of the observed pion-proton system. Technique of data analysis allows only for measuring total asymmetry but study done with the models gives possibility to estimate what is the contribution of emission time difference and pair separation to the total asymmetry. Such study done with Therminator suggests that protons are mainly emitted earlier than pions. In UrQMD such phenomena is not observed. Both models show that in average protons are shifted to the edge of the source.


\begin{thebibliography}{99}
\bibitem{asym:llen}
            \textit{Lednicky R., et. al.} //
            {Phys. Lett.} {B. 1996. B.373. P.30}
\bibitem{asym:vlpx}
            \textit{Voloshin S., et. al} //
            {Phys. Rev. Lett. 1997. V.79. P.4766.}
\bibitem{mzappb}
            \textit{Zawisza M.} //
            {Acta Phys. Pol.} {B. 2009. B40. P.1163}
\bibitem{shzibimike}
            \textit{Chajecki Z., Lisa M.} //
            {Phys. Rev.} {C. 2008. V.78. P.64903}
\bibitem{shkisielbrown}
            \textit{Kisiel A., Brown D.A.} //
            {Phys. Rev.} {C. 2009. V.80. P.64911}
\bibitem{corrfit}
            \textit{Kisiel A.} //
            {Nucleonika} {2004. V.49 (Supl. 2). P.81}
\bibitem{therm}
            \textit{Chojnacki M., et. al.} //
            {These proceedings}
\bibitem{urqmd}
            \textit{Bass S.A. et. al} //
            {Prog. Part. Nucl. Phys} {1998. V.41 P.255} \\
            \textit{Bleicher M., et. al} //
            {J. Phys.} {G. 1999 V.25 P.1859}

\end{thebibliography}
\end{document}